
\input harvmac

\def\N{{\cal N}}


\lref\NekrasovQD{ N.~A.~Nekrasov,
arXiv:hep-th/0206161.
}
\lref\FlumeAZ{ R.~Flume and R.~Poghossian,
arXiv:hep-th/0208176.
}
\lref\BruzzoXF{ U.~Bruzzo, F.~Fucito, J.~F.~Morales and
A.~Tanzini,
arXiv:hep-th/0211108.
}
\lref\DV{ R.~Dijkgraaf and C.~Vafa,
Nucl.\ Phys.\ B {\bf
644}, 3 (2002) [arXiv:hep-th/0206255];
Nucl.\ Phys.\ B {\bf 644}, 21
(2002) [arXiv:hep-th/0207106];
arXiv:hep-th/0208048;
[arXiv:hep-th/0302011].
}
\lref\ovafan{ H.~Ooguri and C.~Vafa,
Nucl.\ Phys.\ B {\bf 641}, 3 (2002)
[arXiv:hep-th/0205297].
}
\lref\work{ M.~Matone and L.~Mazzucato, work in progress.}
\lref\DGKV{ R.~Dijkgraaf, S.~Gukov, V.~A.~Kazakov and C.~Vafa,
[arXiv:hep-th/0210238].
}
\lref\dvsw{ M.~Matone,
Nucl.\ Phys.\ B {\bf 656}, 78 (2003)
[arXiv:hep-th/0212253].
}
\lref\SeibergRS{ N.~Seiberg and E.~Witten,
Nucl.\ Phys.\ B {\bf 426}, 19
(1994) [Erratum-ibid.\ B {\bf 430}, 485 (1994)]
[arXiv:hep-th/9407087].
}
\lref\DijkgraafXD{ R.~Dijkgraaf, M.~T.~Grisaru, C.~S.~Lam, C.~Vafa and D.~Zanon,
[arXiv:hep-th/0211017].
}
\lref\CDSW{ F.~Cachazo, M.~R.~Douglas, N.~Seiberg and E.~Witten,
JHEP {\bf 0212}, 071 (2002) [arXiv:hep-th/0211170].
}
\lref\CachazoZK{ F.~Cachazo, N.~Seiberg and E.~Witten,
JHEP {\bf 0302}, 042 (2003) [arXiv:hep-th/0301006];
JHEP {\bf 0304}, 018 (2003) [arXiv:hep-th/0303207].
}
\lref\WittenYE{E.~Witten,
arXiv:hep-th/0302194.
}
\lref\pestun{A.~Dymarsky and V.~Pestun,
[arXiv:hep-th/0301135].
}
\lref\AganagicXQ{
M.~Aganagic, K.~Intriligator, C.~Vafa and N.~P.~Warner,
arXiv:hep-th/0304271.
}
\lref\rela{ M.~Matone,
Phys.\ Lett.\ B {\bf 357}, 342 (1995)
[arXiv:hep-th/9506102];
Phys.\ Rev.\ D {\bf 53}, 7354 (1996) [arXiv:hep-th/9506181].
}
\lref\proof{ G.~Bonelli, M.~Matone and M.~Tonin,
Phys.\ Rev.\ D {\bf
55}, 6466 (1997) [arXiv:hep-th/9610026].
}
\lref\non{G.~Bonelli and M.~Matone,
Phys.\ Rev.\ Lett.\  {\bf 77}, 4712 (1996) [arXiv:hep-th/9605090];
}
\lref\miro{ L.~Chekhov and A.~Mironov,
Phys.\ Lett.\ B {\bf 552}, 293
(2003) [arXiv:hep-th/0209085].
H.~Itoyama and A.~Morozov,
[arXiv:hep-th/0211245];
}
\lref\BertoldiAB{ G.~Bertoldi,
arXiv:hep-th/0305058.
}
\lref\nacu{ S.~G.~Naculich, H.~J.~Schnitzer and N.~Wyllard,
Nucl.\ Phys.\ B {\bf
651}, 106 (2003) [arXiv:hep-th/0211123];
JHEP {\bf 0301}, 015 (2003)
[arXiv:hep-th/0211254].
}
\lref\KrausJF{ P.~Kraus and M.~Shigemori,
JHEP {\bf 0304}, 052 (2003) [arXiv:hep-th/0303104].
}
\lref\michelefernando{ L.~F.~Alday and M.~Cirafici,
arXiv:hep-th/0304119.
P.~Kraus, A.~V.~Ryzhov and M.~Shigemori,
arXiv:hep-th/0304138.
}
\lref\LG{ M.~Matone,
J.\ Geom.\ Phys.\  {\bf 21}, 381 (1997)
[arXiv:hep-th/9402081].
}
\lref\itoyama{ H.~Itoyama and A.~Morozov,
[arXiv:hep-th/0301136].
}
\lref\sfere{M.~Matone,
JHEP {\bf 0104}, 041 (2001)
[arXiv:hep-th/0103246];
Phys.\ Lett.\  {\bf 514}, 161 (2001)
[arXiv:hep-th/0105041].
}
\lref\KMT{ A.~Klemm, M.~Marino and S.~Theisen,
JHEP {\bf 0303}, 051 (2003) [arXiv:hep-th/0211216].
}
\lref\CachazoJY{ F.~Cachazo, K.~A.~Intriligator and C.~Vafa,
Nucl.\ Phys.\ B {\bf
603}, 3 (2001) [arXiv:hep-th/0103067].
}
\lref\konishione{ K.~Konishi,
Phys.\ Lett.\ B {\bf 135}, 439 (1984).
K.~i.~Konishi and K.~i.~Shizuya,
Nuovo Cim.\ A {\bf 90}, 111 (1985).
}
\lref\VenezianoAH{
G.~Veneziano and S.~Yankielowicz,
Phys.\ Lett.\ B {\bf 113}, 231 (1982).
}
\lref\KovnerIM{
A.~Kovner and M.~A.~Shifman,
Phys.\ Rev.\ D {\bf 56}, 2396 (1997)
[arXiv:hep-th/9702174].
}

\Title{\vbox{\baselineskip11pt\hbox{hep-th/0305225}
\hbox{DFPD/02/TH/36} \hbox{SISSA 23/2003/EP} }} {\vbox{
\centerline{Branched Matrix Models and the Scales}
\smallskip
\centerline{of Supersymmetric Gauge Theories}
\vskip 1pt }}
\smallskip
\centerline{ Marco Matone$^1$ and Luca Mazzucato$^2$}
\bigskip
\centerline{$^1$Dipartimento di Fisica ``G. Galilei'', Istituto
Nazionale di Fisica Nucleare,} \centerline{Universit\`a di Padova,
Via Marzolo, 8 -- 35131 Padova, Italy}
\medskip
\centerline{$^2$International School for Advanced Studies,
Trieste, Italy}
\bigskip
\vskip 0.5cm
 \noindent
In the framework of the matrix model/gauge theory correspondence,
we consider supersymmetric $U(N)$ gauge theory with $U(1)^N$
symmetry breaking pattern. Due to the presence of the
Veneziano--Yankielowicz effective superpotential, in order to
satisfy the $F$--term condition $\sum_iS_i=0$, we are forced to
introduce additional terms in the free energy of the corresponding
matrix model with respect to the usual formulation. This leads to
a matrix model formulation with a cubic potential which is free of
parameters and displays a branched structure. In this way we
naturally solve the usual problem of the identification between
dimensionful and dimensionless quantities. Furthermore, we need
not introduce the ${\cal N}=1$ scale by hand in the matrix model.
These facts are related to remarkable coincidences which arise at
the critical point and lead to a branched bare coupling constant.
The latter plays the role of the ${\cal N}=1$ and ${\cal N}=2$
scales tuning parameter. We then show that a suitable rescaling
leads to the correct identification of the ${\cal N}=2$ variables.
Finally, by means of the mentioned coincidences, we provide a
direct expression for the ${\cal N}=2$ prepotential, including the
gravitational corrections, in terms of the free energy. This
suggests that the matrix model provides a triangulation of the
istanton moduli space.

\vskip 0.5cm

\Date{May 2003}
%
\baselineskip14pt

%
%


\newsec{Introduction}

During the last year, our understanding of the nonperturbative
dynamics of four--dimensional supersymmetric gauge theory has
achieved a dramatic advance. Motivated by insights from the
geometric engineering perspective \CachazoJY\ovafan, in a
series of papers \DV\ Dijkgraaf and Vafa have proposed that some
exact holomorphic quantities of ${\cal N}=1$ supersymmetric gauge
theories are captured by an auxiliary matrix model. In particular,
under the assumption that the low energy $F$--term physics is
described by a glueball superfield, they proposed that the ${\cal
N}=1$ effective superpotential is completely obtained by
evaluating the genus zero free energy of the related matrix model.
This conjecture has been proved by two different techniques, first
by showing that the relevant superspace diagrams reduce to a zero
dimensional theory \DijkgraafXD\ and then by showing that the
generalized Konishi anomaly equations in the chiral ring of the
gauge theory are equivalent to the loop equations of the related
matrix model \konishione\CDSW\CachazoZK\WittenYE.
Very recently, an apparent discrepancy has been found
between the standard field theory computation and the matrix model
result, both in the perturbative approach \KrausJF\ and from the
Konishi anomaly point of view \michelefernando. The solution
of this puzzle has been proposed in \AganagicXQ\ by investigating the ambiguity
in the UV completion of the supersymmetric gauge theories.

A strong check of the DV conjecture can be performed in the cases
where the gauge theory exact quantities are already available,
e.g. by testing the well known ${\cal N}=2$ supersymmetric gauge
theory, namely Seiberg--Witten theory \SeibergRS. A recent step in
this direction \dvsw\ concerns the exploration of the way in which
the well known duality structure of the SW theory appears
inside the matrix model itself (see also \miro\nacu\pestun).
The usual matrix model formulation of the SW theory \DGKV\ poses
a number of questions. Let us briefly discuss the main issues.

The first concerns the extremization
of the superpotential. The matrix model conjecture
for the SW theory with $U(N)$ gauge group requires the soft
breaking of ${\cal N}=2$ to ${\cal N}=1$ by adding a tree level superpotential.
Now, due to the usual structure of the
Veneziano--Yankielovicz effective superpotential \VenezianoAH,
namely the appearance of the log terms, the $F$--term
condition for the glueball superfields $S_i$, strictly speaking,
cannot give the expected extremum
condition $\sum_{i=1}^N S_i=0$.

A second crucial point is related to the fact that the most basic
feature of the SW gauge theory, namely the duality structure,
is not displayed in its matrix model counterpart. As explained
in \dvsw, the first step in investigating such deep aspect
is to consider the scaling properties of the matrix model free
energy. In particular, it was shown that the natural variable
in order to display the SW duality is a rescaled version of the
glueball superfield. Therefore, the question arise whether
there exists a formulation of the conjecture that
by itself provides this additional structure.

Another important issue is related to the introduction of
dimensionful quantities in the matrix model. On one hand,
in the usual formulation one introduces by hand a cutoff $\Lambda$
directly in the free energy of the matrix model, relying on
a gauge theory expectation. The presence of such a dimensionful
object in the matrix model provides in turn serious troubles
in analyzing the monodromy properties of the free energy.
Moreover, the identification of the
$u$-modulus of the SW theory and of the dynamically generated
scale of the gauge theory is performed at the critical point
by comparing it to the known SW curve. In this respect,
one might ask if any information about the scales
and the modulus of the gauge theory is present even outside
the critical point. Finally, in the usual approach one identifies
the dimensionless 't Hooft couplings of the matrix model
with the dimensionful gauge theory glueball superfields. This
identification then leads to some problems in the interpretation
of the extremum condition, namely introducing the
concept of ``eigenvalue holes'' corresponding to the unstable
extrema of the matrix model potential.
Actually, all the issues considered so far will turn out
to be tightly related one to each other.

In this paper we will present
a first investigation of this fact, mainly stating the most
interesting results. The details of the calculations and
crucial generalizations will be given in \work.
In Section 2, by addressing the problem of the minimization
of the superpotential, we will show that
$$
{\cal F}_{0}(-S_2,-S_1)\neq{\cal F}_{0}(S_1,S_2).
$$
Requiring that this symmetry exactly holds, one should modify
the free energy by adding some bilinear terms such that the
new free energy
$$
{\cal F}_0^{(k)}= {\cal F}_0+
\delta{\cal F}_{0}^{(k)},
$$
displays different branches that depend on the odd number $2k+1$.
In this way we obtain
$$
{\cal
F}_{0}^{(k)}(e^{i \kappa}S_2,e^{-i \kappa}S_1)={\cal
F}_{0}^{(k)}(S_1,S_2),
$$
where $\kappa\equiv (2k+1)\pi$. Then we will show that, in order to compare the matrix model
quantities with the well known SW
exact results, we have to perform a rescaling transformation
on the matrix model variables \dvsw.
In Section 3 we
show that the proposed free energy is given by the matrix model
with potential
$$
W^{(k)}(\Phi)=\Tr\,\Bigl({1\over2}e^{{i\over 2}\kappa}\Phi^2+
{1\over3}e^{{3i\over 4}\kappa}\Phi^3-{{\bf 1}\over12}\Bigr),
$$
where, with respect to the usual formulation,
the couplings disappear. A crucial term in the evaluation of the
matrix model is the gaussian contribution which, due to the phases,
will be given by
$$
e^{-{i\over4}\kappa(M_1^2-M_2^2)}.
$$
This will also solve the questions related
to the identification of dimensionful quantities.
We then show that, in order to reproduce
the expected extremum condition, the correct gauge theory bare coupling is
$$
\tau_0\longrightarrow
\tau_0^{(k)}={2\over\pi i}
\ln{\Lambda_2\over\Lambda_1}-{\kappa\over2\pi},
$$
where $\Lambda_1$ and $\Lambda_2$ are the dynamically generated scale of
the ${\cal N}=1$ and ${\cal N}=2$ supersymmetric gauge theories, respectively.
This clearly shows the role of the bare coupling as the ${\cal N}=1$ and ${\cal N}=2$
scales tuning parameter.
Moreover, the nonperturbative relation found in \dvsw\ implies that at
the critical point
$$
\Lambda_1^2=4u,
$$
where $u$ is the SW modulus. Finally, it turns out that
the free energy directly evaluated at the critical point
is equal, up to a numerical factor, to the one obtained by
integrating twice the effective coupling constant
evaluated at the extremum.
In other words, we find that
$$
e^{{\hat{\cal F}/\Lambda_2^6}}={1\over{\rm
Vol}(U(M_1))\times{\rm Vol}(U(M_2))}{\int {\cal D}\Phi_1{\cal
D}\Phi_2 e^{-W^{(k)}(\Phi_1,\Phi_2)}}|_{M_1=e^{i\kappa}M_2\equiv {\cal
S}/\Lambda_2^3}.
$$
This in turn provides a direct matrix model formulation for the
${\cal N}=2$ instantons, including the gravitational corrections,
and shows that the matrix model provides a kind of triangulation
of the instanton moduli space.

\newsec{Branches from the Symmetry of the Free Energy}

In this section we start investigating the structure of the planar
contribution to the free energy ${\cal F}_0$ of the matrix model.
Since the matrix model formulation reproducing the CIV
prepotential \CachazoJY\ has the advantage of a direct test with
the established results of SW theory, we focus on this case.
Nevertheless, many of the results we will find have a more general
validity as they concern the matrix model formulation by itself
rather than the specific SW realization.

Let us then start with the free energy derived in \DGKV. The
strategy will be to analyze its structure in relation with the
exact ${\cal N}=2$ results it should reproduce at the extremum. In
the analysis we will use the relationship between the ${\cal
N}=2$ $u$--modulus and the prepotential for the gaugino condensate
derived in \dvsw. This investigation will lead to introduce
additional terms to ${\cal F}_0$ that
will help us in deriving the matrix model formulation.

\subsec{Stating the problem}

In our explicit computation we will consider the
case of a $U(2)$ gauge group spontaneously broken to $U(1)\times
U(1)$ with a cubic tree level superpotential. In this simple case
we have two superfields $S_1$ and $S_2$, that will describe the effective
Abelian dynamics. Let us consider the matrix model and write down
the expression of the free energy \DGKV\
\eqn\dacompararesotto{
{\cal F}_{0}(S_i)=\sum_{j=1,2}{S_j^2\over2} \ln {S_j \over
\Delta^3}-(S_1+S_2)^2 \ln {\Lambda \over
\Delta}+\Delta^6\sum_{n\geq3}\sum_{j=0}^n c_{n,j}
\Bigl({S_1\over\Delta^3}\Bigr)^{n-j}\Bigl({S_2\over\Delta^3}\Bigr)^j.
}
It turns out that the coefficients of the expansion satisfy the
property
\eqn\gaiardi{ c_{n,j}=(-1)^nc_{n,n-j},\qquad
c_{n,j}=(-1)^j|c_{n,j}|.}
Eq.\dacompararesotto\ has been
derived in \DGKV\ by the matrix model formulation, except for the
term depending on $\Lambda$ that should be added by hand, as
expected from the gauge theory. This expression for
${\cal F}_{0}$ differs by the relative sign between the infinite
sum and the first two contributions with respect to \DGKV\ (as we
will see, this fits with the implied expressions for the ${\cal
N}=2$ modulus $u$ and the effective coupling constant $\tau$).
Note that besides the second term, also the third one, as follows
by $c_{n,j}=(-1)^nc_{n,n-j}$, is symmetric under the
transformation $S_1\to-S_2$, $S_2\to-S_1$. However, due to the
log term, we have
$$
{\cal F}_{0}(-S_2,-S_1)\neq{\cal F}_{0}(S_1,S_2).
$$
On the other hand, to get the value
$\langle S_1\rangle=-\langle S_2\rangle\equiv S$ at the extremum
we need the exact symmetry. The extremum corresponds to the
minimum of $W_{eff}$,
which is usually evaluated by setting the bare coupling constant
$\tau_0$ to zero. This gives
\eqn\cond{ \sum_{i}\tau_{ij}=0,\qquad j=1,2, }
where $\tau_{ij}={\partial^2{\cal F}_0\over\partial S_i\partial S_j}$,
that is $\tau_{11}=\tau_{22}=-\tau_{12}$.
On the other hand, setting $S_1=-S_2$, in $\tau_{ij}$, gives
\eqn\odnfs{ \tau_{11}-\tau_{22}=\ln\Bigl({S_1\over\Delta^3}\Bigr)-
\ln\Bigl(-{S_1\over\Delta^3}\Bigr)=(2k+1)\pi i,}
$k\in{\bf Z}$, so that $\tau_{11}-\tau_{22}$
does not vanish. Therefore, the symmetry of ${\cal F}_0$ under $S_1\to-S_2$,
$S_2\to-S_1$ should be exact in order to get the critical value
$S_2=-S_1$. This suggests modifying ${\cal F}_{0}$ in such a way
that the following two crucial features hold:

\item{(1)} The critical value for $S$ as a function of
$\Delta$ and $\Lambda$, which follows from the condition
$\tau_{12}+\tau_{11}=0$ evaluated at the extremum, be unchanged and fit with the exact result
\refs{\DGKV,\dvsw}.

\item{(2)} The effective gauge coupling constant $\tau_{11}$ evaluated at the critical point
reproduces the well known SW exact result \SeibergRS.

It turns out that if one chooses a vanishing bare coupling constant, then
there is a modification to the free energy satisfying the above conditions, except for an
apparently irrelevant term. The additional term reads
\eqn\totale{\delta{\cal F}_{0}^{(k)}=
{\Delta^3\over12}(S_1-S_2)-{i\over
4}\kappa(S_1^2-S_2^2)+{i\over2}\kappa S_1S_2
-{3\over4}(S_1^2+S_2^2), }
where
$$\kappa\equiv (2k+1)\pi,\qquad k\in{\bf Z},$$
so that the modified free energy ${\cal F}_0^{(k)}\equiv {\cal F}_0+
\delta{\cal F}_{0}^{(k)}$
displays the requested symmetry \eqn\simmetria{ {\cal
F}_{0}^{(k)}(e^{i \kappa}S_2,e^{-i \kappa}S_1)={\cal
F}_{0}^{(k)}(S_1,S_2). }
In the following, after discussing the crucial scaling properties of the free energy,
we will check that the addition of \totale\ to the free
energy reproduces the requested features at the extremum (see also \work).
However, we will see that it remains a ``minor discrepancies''. Removing
it will lead to the exact formulation with a unequivocally fixed bare coupling constant.

\subsec{Rescaling the free energy}

In \dvsw\ it has been shown that the free energy satisfies a
scaling property which selects the natural variables to make
duality transparent. In this respect, we note that the duality one
obtains in ${\cal N}=1$ is the one induced, by consistency, by the
$\Gamma(2)$ monodromy of ${\cal N}=2$.
The scaling property of the free energy is obtained by first
rescaling $S_i$, $\Delta$ and $\Lambda$ \eqn\impor{
S_i\longrightarrow {\cal S}_i=\Bigl({\Lambda\over \Delta}\Bigr)^3
S_i, \qquad \Delta\longrightarrow {\Lambda\over \Delta}
\Delta=\Lambda, \qquad \Lambda\longrightarrow {\Lambda\over
\Delta}\Lambda= {\Lambda^2\over \Delta}, } and then performing the
map \eqn\oknco{{\cal F}_{0}^{(k)}(S_i,\Delta,\Lambda)
\longrightarrow {\cal F}_{0}^{(k)}({\cal
S}_i,\Lambda,\mu\Lambda)=\mu^6{\cal
F}_{0}^{(k)}(S_i,\Delta,\Lambda),} where
$\mu\equiv{\Lambda\over \Delta}$.
Note that since the comparison with the SW curve gives \DGKV\
$\Delta^2=4u$, we thus have
$\mu=({\Lambda^2/4u})^{1/2}$.
We observe that whereas in the
original free energy the scale $\Lambda$ appears in pair with
$\Delta$, in the rescaled free energy we have that $\mu$ is
``decoupled" from $\Lambda$. More precisely, ${\cal F}_0^{(k)}$
has the structure \eqn\additive{\Lambda^{-6}{\cal
F}_{0}^{(k)}({\cal S}_i,\Lambda,\mu\Lambda)= {\cal
H}^{(k)}\Bigl({{\cal S}_1\over\Lambda^3},{{\cal
S}_2\over\Lambda^3}\Bigr)- \Bigl({{\cal S}_1\over\Lambda^3}+{{\cal
S}_2\over\Lambda^3}\Bigr)^2\ln\mu.}

Let us show that the dependence of $S$ on
$\Lambda$ and $\Delta$ still follows after modifying the free energy as in
\totale. The extremum condition \cond\ holds unchanged also
for the rescaled variables, in particular
${\cal S}\equiv {\cal S}_1=e^{i \kappa}{\cal S}_2$.
Set $\tau^{(k)}={\partial^2 {\cal F}_0^{(k)}\over\partial S_i\partial S_j}$. From
the condition $\tau_{11}^{(k)}+\tau_{12}^{(k)}=0$ we get the
expansion of $\mu$ \eqn\oifw{ \mu^4={{\cal
S}\over\Lambda^3}\exp\Bigl[\sum_{n\geq3}b_n \Bigl({{\cal
S}\over\Lambda^3}\Bigr)^{n-2}\Bigr], } where
$$
b_n=\sum_{j=0}^n|c_{n,j}|(n-j)(n-2j-1).
$$
We shall see that this
expansion leads to a set of relations that constrain the coefficients
of the free energy. Inverting \oifw\ as a
series for ${\cal S}/\Lambda^3$ in powers of $\mu^4$, one obtains
\eqn\laserie{ {\cal
S}=\Lambda^3(\mu^4+6\mu^8+140\mu^{12}+4620\mu^{16}+\ldots), }
that
expressed in terms of $S$ coincides with the series given in
\DGKV.

\subsec{An unwanted term}

The asymptotic expansion for
$\tau^{(k)}\equiv\tau_{11}^{(k)}$ reads\foot{In the following
expressions we have rescaled
$\tau^{(k)}$ by ${1/\pi i}$.}
\eqn\taucappa{
\tau^{(k)}=-{\kappa\over2\pi}+{1\over2\pi i}\ln{{\cal
S}\over\Lambda^3} +{1\over4\pi
i}\sum_{n\geq3}n(n-1)a_n\Bigl({{\cal
S}\over\Lambda^3}\Bigr)^{n-2}, }
where $a_n=\sum_{j=0}^n|c_{n,j}|$.
As a check of the formulation outlined so far, we would like to
compare the result of the matrix model computation \taucappa\  to
the known expression of the SW effective coupling constant $\tau$.

In order to do this we have to plug the expansion
\laserie\ into \taucappa.
Since $\mu^4(u)=2^{-6}(\Lambda_{SW}^2/ u)^2$, where $\Lambda_{SW}=\sqrt 2\Lambda$,
by using the asymptotic expansion of $u(a)$ in \SeibergRS\ we find
\eqn\taumm{\eqalign{
\tau^{(k)}=&-{\kappa\over2\pi} +\tau_{SW},} }
where the well known expression for the SW gauge coupling \SeibergRS\ reads,
after setting $\hat a=a/\Lambda_{SW}$,
\eqn\tausw{
\tau_{SW}={2i\over\pi}\ln2+{2i\over
\pi}\ln{\hat a}+ {3\over4\pi
i}\hat a^{-4}+ {105\over2^7\pi
i}\hat a^{-8} +{165\over2^7\pi
i}\hat a^{-12}+\ldots }
Notice that the term $-{\kappa\over4\pi}{\cal S}^2$
in the onshell rescaled free energy, which generates the discrepancy \taumm,
cannot be reabsorbed by changing the phase of ${\cal
S}$. Actually, the only phase that leaves the perturbative series
of the onshell free energy
$\sum_{n\geq3}a_n({\cal S}/\Lambda^3)^n$ invariant is
$e^{2l\pi i}$, $l\in{\bf Z}$. On the other hand, we have
$$
{\cal F}_0^{(k)}(e^{2l\pi i}{\cal S})={\cal F}_0^{(k-l)}({\cal
S}),
$$
so that a term ${\cal S}^2$ multiplied by a half odd number would
survive. The fact that $\tau^{(k)}$ does not exactly coincide with
the SW effective coupling constant is a crucial question.
Understanding and removing this discrepancy is a key step in our
investigation.

\subsec{Coincidences at the extremum}

By evaluating the relevant quantities at the extremum, some
interesting coincidences arise.
The first step is a remark that, although obvious, needs to be
stressed. This concerns how the prepotential is evaluated at the
extremum. As we said, one first evaluates\foot{In this subsection we
omit the superscript $k$ labelling the branches.} $\tau$ at the
extremum, then integrates it twice with respect to ${\cal S}$ (with
care on the integration constants). Of course, this should be
different from the function one obtains by directly
evaluating it, that is ${\cal F}_0({\cal S}_i,\Lambda, \mu\Lambda)$
with ${\cal S}_i$ and $\mu^4$ replaced by their expressions at the
extremum. We denote this function as
$$
\hat {\cal F}_0({\cal S})\equiv{\cal
F}_0|_{{\cal S},\mu}
$$
where here and in the
following we use the notation
$$
f|_{\cal S}\equiv f|_{{\cal S}\equiv {\cal S}_1=e^{i \kappa}{\cal
S}_2},\qquad f|_{{\cal S},\mu}\equiv f|_{{\cal S}\equiv {\cal
S}_1=e^{i \kappa}{\cal S}_2,\mu=\mu({\cal S}).}
$$
Remarkably, it turns out that directly evaluating ${\cal F}_0$ at the extremum
one gets
\eqn\oqaidxc{\hat {\cal F}_0({\cal S})=4 {\cal F}_0(\cal{S}).}
Let us set $\hat {\cal S}_D^i({\cal S})\equiv {\partial
{\cal F}_0\over\partial {\cal S}_i}|_{{\cal S},\mu}$.
We have
\eqn\oqaidxcbbb{
\hat {\cal
S}_D^2({\cal S})=-\hat {\cal S}_D^1({\cal S})=-2 {\cal S}_D({\cal
S})=-{1\over2}{\partial\hat{\cal F}_0 \over\partial{\cal S}},}
where ${\cal S}_D({\cal S})= {\partial {\cal F}_0({\cal S})\over
\partial{\cal S}}$.
Since $\mu$ appears in ${\cal F}_0$ only through the
term $-({\cal S}_1+{\cal S}_2)^2\ln\mu$, it follows that in
evaluating $\hat {\cal F}_0$ and $\hat {\cal S}_D^i$ we do not
need the value of $\mu$ at the extremum (given in Eq.\oifw). In
other words, just setting ${\cal S}\equiv{\cal S}_1=e^{i
\kappa}{\cal S}_2$, we obtain both $\hat {\cal F}_0$ and $\hat
{\cal S}_D^i$. In order to evaluate ${\cal F}_0$ directly at the
extremum we need only this ``trivial part'' of the condition coming from
the extremum.
In particular, we have
$$
\tau=
{1\over2}{\partial\hat{\cal S}_D\over\partial{\cal S}}
={1\over4}{\partial^2\hat{\cal F}_0\over\partial{\cal S}^2}.
$$

Consider now the following nonperturbative
relation \dvsw\
\eqn\muuu{ \mu^4={3\cdot 2^4\pi
i\over\Lambda^6}\Bigl({\cal F}_0^{(k)}-{{\cal S}\over 2}{\partial
{\cal F}_0^{(k)}\over\partial {\cal S}}\Bigr), }
which is the analogue of the $U(1)_R$ anomaly equation derived in
SW theory \rela.
A first interesting consequence of
the above coincidences is that this relation between $\mu$ and the
prepotential also holds, except for a factor $4$, if one first
computes the Legendre transform of ${\cal F}_0$ with respect to
${\cal S}_i^2$, and then evaluates it at the extremum.
Since, as
we said, the critical values are independent of the value of $\mu$
at the extremum, by \muuu\ and \oqaidxc\ we obtain
\eqn\muuubis{{\Lambda^6\over6\pi i}\mu^4= (2{\cal F}_0-{\cal S}_j
\partial_{{\cal S}_j} {\cal F}_0
)|_{\cal S} =8{\cal F}_0-4{\cal S}
\partial_{\cal S}{\cal F}_0
=2\hat{\cal F}_0-\hat{\cal S}_j\hat{\cal S}_{D}^{j} =2\hat{\cal
F}_0-\hat{\cal S}\partial_{\cal S}\hat{\cal F}_0, } where
$\hat{\cal S}_1\equiv{\cal S}$, $\hat{\cal S}_2\equiv-{\cal S}$.
Among the various
versions \muuubis\ of the relation \muuu, there is only one which
can be satisfied by the unrescaled ${\cal F}_0(S_i)$, i.e.
\eqn\muuubisancora{u={3\pi i\over2\Lambda^4}\Bigl(2{\cal
F}_0(S_i)-S_j
\partial_{S_j}{\cal
F}_0(S_i) \Bigr)|_{S,\mu}, }
where $\mu=(\Lambda^2/4u)^{1/2}$.
This is the version of the relation found in \dvsw\ in the form derived
by Dymarsky and Pestun \pestun\ (see also \itoyama). In this
respect we note that while the relation between $\mu$ and ${\cal
F}_0({\cal S}_i,\Lambda,\mu\Lambda)$ holds in the versions given
in \muuubis, this is not the case for ${\cal
F}_0(S_i,\Delta,\Lambda)$ that satisfies the relation only in the
case in which the extremum is considered after the Legendre
transform with respect to $S_i^2$ has been evaluated, that is
Eq.\muuubisancora.

The detailed analysis of these coincidences
will be presented elsewhere \work.
The origin of the observed coincidences relies on two crucial facts,
namely the symmetry \simmetria\ of the
free energy and the remarkable structure \additive, that emerges after the rescaling.
Moreover, due to the latter structure, one can easily prove that the gauge
coupling is not affected by replacing the term $\ln\mu$ by
a generic function $f(\mu)$, including $f\equiv0$ (in this case $\tau_{ij}={\cal
H}_{ij}$).

\subsec{Recursion relations}

The relation between the ${\cal N}=2$ $u$--modulus and the prepotential, in the context
of the SW theory,
leads to the proof of the SW conjecture \proof. The analogous relation
in the matrix model is given by \muuu. Since this has a nonperturbative nature,
it can then be argued that this relation puts strong constraints on the structure of the
matrix model formulation itself. Remarkably, this is indeed the case as by
\oifw\ and \muuu\ we get
\eqn\edajjie{
\exp\Bigl[\sum_{n\geq3}b_n \Bigl({{\cal
S}\over\Lambda^3}\Bigr)^{n-2}\Bigr]=
1-6{{\cal
S}\over\Lambda^3}+6\sum_{n\geq3}(2-n)a_n\Bigl({{\cal
S}\over \Lambda^3}\Bigr)^{n-1},}
which provides infinitely many conditions on the coefficients $c_{n,i}$ of the ${\cal N}=1$
free energy. Even if apparently these conditions do
not unequivocally fix the $c_{n,i}$ it is plausible
that there exists a simple argument leading to fix them completely.
In particular,
$$
b_{n+3}-{6n\over n+1}b_{n+2}+6na_{n+2}-{6\over n+1}\sum_{j=0}^{n-2}(j+1)(n-j-1)a_{n-j+1}b_{j+3}=0.
$$
which has been explicitly checked up to $n=7$ \work.

\newsec{Branching the Matrix Model}

At this stage it is useful to summarize some questions one
meets in the matrix model formulation of supersymmetric gauge
theories. Even if we are considering the specific case of the
CIV free energy \CachazoJY, the issues
we are dealing with extend to more general cases.

The first problem concerns the gauge coupling constant.
A starting point of our investigation was \odnfs\ showing that
$S_1=-S_2$ is not a solution. On the other hand, this is related
to the lack of symmetry of the original free energy
\dacompararesotto\ under $S_1\to -S_2$, $S_2\to -S_1$. We then saw
that this symmetry, and therefore the solution $S_1=e^{i
\kappa}S_2$, can be restored by including additional terms to the
free energy \dacompararesotto\ depending on the odd number
${\kappa\over\pi}=2k+1$ which specifies the symmetry, namely
$$S_1\longrightarrow e^{i \kappa}S_2,\qquad S_2\longrightarrow e^{-i \kappa}S_1.$$
In this way one obtains the correct critical values for $\mu$, ${\cal
F}_0^{(k)}$ and therefore $\tau^{(k)}$. However, as we said,
in comparing $\tau^{(k)}$ with the SW effective coupling
constant, one sees that they coincide up to the term
$-{\kappa\over2\pi}$. This is not a minor question. First of all
note that the exact expression of $\tau$ is necessary to get the
correct monodromy. In general, rescaling or adding a constant to a
function breaks its M\"obius polymorphicity, in our case
\eqn\mobbius{ \tau \longrightarrow
\tilde\tau={A\tau+B\over C\tau+D},}
where the constants are the
entries of the matrices in $\Gamma(2)$. The only possibility to
add a constant by preserving the monodromy properties of $\tau$ is
that such a constant corresponds to a translation in $\Gamma(2)$.
On the other hand, $\tau^{(k)}$ in \taucappa\ differs from the
SW effective coupling constant $\tau$ by the constant
$-{\kappa\over2\pi}$, that is \eqn\eccellente{
\tau^{(k)}=\tau-{\kappa\over2\pi}, } and since the difference
$\tau^{(k)}-\tau$ is a non integer number, \eccellente\ cannot
correspond to a $\Gamma(2)$ monodromy of $\tau$.

Another open question is related to
the fact this formulation of the conjecture provides an expression for ${\cal
F}_0^{(k)}(S_j,\Delta,\Lambda)$ while, in order to display
the duality structure, we are forced to consider
its rescaled version ${\cal F}_0^{(k)}({\cal
S}_j,\Lambda,\mu\Lambda)= \mu^6{\cal
F}_0^{(k)}(S_j,\Delta,\Lambda)$, as observed in \dvsw. The properties of ${\cal
F}_0^{(k)}({\cal S}_j,\Lambda,\mu\Lambda)$ indicate that this
rescaling actually hides a property of the matrix model
formulation which is still to be understood.

The last question concerns the nature of $\Delta$, which
is related to the scaling properties of the free energy.
In the usual formulation, $\Delta$ is to be identified
with $2\sqrt u$ by comparing the matrix model curve at the
extremum with the SW curve. But one should be led to investigate
the meaning of $\Delta$ even outside the critical point. On the other hand,
in passing to the effective superpotential of the gauge theory
we would like to consider the ${\cal N}=1$ dinamically generated scale
$\Lambda_1$ rather than $\Delta$.
Related to these questions is the unpleasant feature that, in
order to derive the free energy, one is forced
to identify dimensionless quantities with dimensionful ones.

The above list concerns the main, strictly connected, questions
related to the matrix model formulation of supersymmetric gauge
theories. The problem is to understand whether there exists an
exact matrix model formulation free of the above
problems. In the following we will see that in fact
there exists such a formulation and that, while possessing some of
the relevant features of the original formulation, it leads
to a natural explanation based on the two
different dynamically generated scales of the ${\cal N}=1$ and
${\cal N}=2$ theories.

\subsec{Branches in the matrix model}

Let us consider as our starting point
the matrix model with cubic potential \DGKV\KMT.
We now show that a suitable modification of that model actually
leads to a formulation free of the problems outlined in the previous
section. In order to
find the matrix model potential we first note that we should
introduce, in the usual cubic potential,
the branches we labelled by the integer $k$
$$
W^{(k)}=\Tr\,(m^{(k)}\Phi^2+g^{(k)}\Phi^3), \qquad k\in{\bf Z}.
$$
To specify the meaning of the index in the matrix potential we
try to eliminate some of the dimensional problems one has from
the very beginning in the formulation. First, since $\Phi$
are dimensionless quantities, in order
to be consistent one should require that $m^{(k)}$ and
$g^{(k)}$ be dimensionless quantities.

Eliminating dimensionful quantities from the potential
leads us to consider the following dimensionless branched matrix
model potential \eqn\clamoroso{
W^{(k)}(\Phi)=\Tr\,\Bigl({1\over2}e^{{i\over 2}\kappa}\Phi^2+
{1\over3}e^{{3i\over 4}\kappa}\Phi^3-{{\bf 1}\over12}\Bigr),}
$\kappa\equiv (2k+1)\pi, k\in{\bf Z}$, where we added an
integration constant for future purpose. With respect to the previous
formulation this potential does not contain any parameter. Even if
surprising, we will show that this will reproduce the exact ${\cal N}=2$
results and will be free of the problems outlined so far. Note that the
$\kappa$--dependence can be completely absorbed in the matrix
redefinition $\Psi_k=e^{{i\over 4}\kappa}\Phi$ so that
\eqn\clamorosoooo{ W^{(k)}(\Phi)= \Tr\,\Bigl({1\over2}\Psi_k^2+
{1\over3}\Psi_k^3-{{\bf 1}\over12}\Bigr),}
in other words
$$
W^{(k)}(\Phi)=W^{(0)}(\Psi_k).
$$
We consider the two cut
solution in which $M_1$ eigenvalues fluctuate around to the
critical point $a_1=0$ and $M_2=M-M_1$ eigenvalues
fluctuate around the other critical point
$a_2=-e^{-{i\over4}\kappa}$. As usual, one passes to the eigenvalue
representation getting as Jacobian the square of the Vandermonde
determinant.
In terms of the fluctuations around the two vacua
$$
\lambda_i=a_1+\nu_{1i}, \quad i=1,\ldots,M_1,\qquad
\lambda_i=a_2+\nu_{2i}, \quad i=M_1+1,\ldots,M,
$$
we can exponentiate the Vandermonde determinant and
obtain the matrix model around this vacuum
\eqn\zetacappa{ Z^{(k)}={1\over{\rm Vol}(U(M_1))\times{\rm
Vol}(U(M_2))} \int D\Phi_1D\Phi_2{\rm
e}^{-W_1^{(k)}(\Phi_1)-W_2^{(k)}(\Phi_2)-W_I^{(k)}
(\Phi_1,\Phi_2)},}
where
$$
W_1^{(k)}=\Tr\,\Bigl({1\over 2}e^{{i\over 2}\kappa}\Phi_1^2+
{1\over3}e^{{3i\over 4}\kappa}\Phi_1^3-{{\bf 1}\over12}\Bigr),
$$
$$
W_2^{(k)}=-\Tr\,\Bigl({1\over 2}e^{{i\over 2}\kappa}\Phi_2^2
-{1\over 3}e^{{3i\over 4}\kappa}\Phi_2^3-{{\bf 1}\over12}\Bigr).
$$
Note that the constant term $-{\bf 1}/12$ in \clamoroso\ is the
one which leads to a constant contribution to
$W_1^{(k)}+W_2^{(k)}$ that vanishes when $M_1=M_2$.
The interaction term $W_I^{(k)}$ is obtained by expanding the log
when exponentiating the Vandermonde determinant.
While referring to \work\ for the details about the evaluation of this
matrix model, here we just comment on the quadratic contribution.
This is important as
it shows that whereas, as in the previous approaches, the propagator
has the ``wrong" sign, this is precisely what we need. While this is
usually seen as a problem of the formulation and so its effect is essentially ignored,
we see that the minus sign leads to the correct expression for the free energy.
If $m$ denotes the coefficient of the quadratic contribution,
then it is usually assumed that this leads to $m^{-{1\over2}(M_1^2+M_2^2)}$.
However, in our case $m=e^{{i\over2}\kappa}$ and
the minus sign for the quadratic contribution to the second matrix potential corresponds
to a minus sign of the exponent
$$
-e^{{i\over2}\kappa}=e^{-{i\over2}\kappa}.
$$
It follows that
the quadratic terms give the following contribution to $Z^{(k)}$
$$
e^{-{i\over4}\kappa(M_1^2-M_2^2)},
$$
which is exactly what we need for reproducing the second term in \totale.
So, we see that the minus sign turns out to be correct in the
matrix model formulation.
Finally, the planar contribution to the free energy reads
$$
 F_0^{(k)}(M_j)=
{1\over12}(M_1-M_2)+\sum_{i=1,2}{M_i^2\over2}\ln
M_i-{i\over4}\kappa(M_1^2-M_2^2)
$$
\eqn\preliminare{ -{i\over 2}\kappa M_1M_2- {3\over4}(M_1^2+M_2^2)
+\sum_{n\geq3}\sum_{j=0}^n c_{n,j}M_1^{n-j}M_2^j. }

\subsec{The gauge theory coupling}

The above results would suggest
identifying $M_i$ with $S_i/\Delta^3$. However, in this case the
new expression \preliminare, besides a global rescaling,
displays two basic differences with respect to
the old free energy \totale.

First of all we note the absence of the term
\eqn\nuovad{ (S_1+S_2)^2 \ln \Bigl( {\Lambda
\over \Delta} \Bigr). }
This term is problematic because, on the one hand, there is no reason for the appearence of
$\Lambda$ in the matrix model free energy; as we already pointed out,
the $\Lambda$ dependence is usually added by hand from a gauge theory guess.
On the other hand, we have seen that this
term plays a basic role in evaluating the extremum of the
effective superpotential.

The second difference is that whereas
in ${\cal F}_0^{(k)}={\cal F}_0+\delta{\cal F}_0^{(k)}$, as in \totale, we have
the term
\eqn\nuovac{ {i\over2}\kappa S_1S_2, }
the analogous contribution in \preliminare, that is $-{i\over2}\kappa M_1 M_2$,
has opposite sign.
Fortunately, we saw that directly evaluating the rescaled
free energy ${\cal F}_0^{(k)}({\cal S}_1,{\cal S}_2)$
at ${\cal S}_1=e^{i \kappa}{\cal S}_2$,
reproduces the SW prepotential except for the term $-(2k+1){\cal
S}^2$. It is precisely because of this change of sign of the term \nuovac\
in \preliminare, that this unwanted additional term is now missing,
so that the first condition coming from the extremum, i.e.
$\tau^{(k)}_{11}=\tau^{(k)}_{22}$, by itself reproduces the
correct SW coupling as in \tausw. But we should now
take into account both the change
of sign of \nuovac\ and simultaneously get the exact expression
for $\mu$. Nevertheless, requiring that the second condition be consistent with the first
one, leads to a new view on the structure of the bare coupling
constant $\tau_0$! The idea is quite natural. Namely, note that in
evaluating the extremum \cond\ we have forgotten the bare coupling
constant by simply putting it to zero. But we can
just use a nonzero value of $\tau_0$ to get the correct critical
values. More precisely, we set
\eqn\nuovaa{ \tau_0\longrightarrow
\tau_0^{(k)}={2\over\pi i}
\ln{\Lambda_2\over\Lambda_1}-{\kappa\over2\pi}. }
Before identifying the two scales $\Lambda_1$ and $\Lambda_2$, we make a
couple of comments on this proposal. A first interesting
consequence of \nuovaa\ is that
\eqn\fondamentale{
\tau_0^{(j)}-\tau_0^{(k)}=k-j, }
that will be discussed in \work. The second observation is that, as we will see,
the term ${2\over\pi i} \ln{\Lambda_2\over\Lambda_1}$ inside
$\tau_0^{(k)}$ compensates the fact that \nuovad\ is now missing,
as it should, since it cannot derive from the matrix model,
from the new free energy \preliminare. Furthermore,
the role of the term ${\kappa\over2\pi}$ in $\tau_0^{(k)}$ is that
of compensating the change of sign of the term \nuovac\ in
\preliminare, a request coming
from the need of obtaining the correct expression for $\mu$ as
given in \oifw.

Let us now identify the scales $\Lambda_1$ and $\Lambda_2$. The
meaning of $\Lambda_2$ is obvious, as it plays the role of the
scale $\Lambda$ appearing in the expression of the ${\cal N}=2$
effective coupling, so that $\Lambda_2\equiv 2^{-1/2}\Lambda_{SW}$.
Since in this new approach the $\Delta$
parameter simply disappeared, the natural choice for $\Lambda_1$ is just the
${\cal N}=1$ dynamically generated scale, as it should appear in
the expression of the effective potential. Therefore, we have
$$
\Lambda_1\equiv\Lambda_{{\cal N}=1},\qquad
\Lambda_2\equiv\Lambda_{{\cal N}=2}.
$$

\subsec{The prescription}

We now consider the link between the matrix model and the gauge theory.
The prescription is to make the following dimensionless identification in
\preliminare\ \eqn\ns{ M_1={S_1\over\Lambda_1^3},\qquad
M_2={S_2\over\Lambda_1^3}, }
with the free energy given by
\eqn\lasoluzione{ {\cal F}_{0}^{(k)}(
S_i,\Lambda_1)=\Lambda_1^6F_{0}^{(k)}\Bigl({S_i\over\Lambda_1^3}\Bigr),
}
that is
\eqn\dacompararesoprabitod{\eqalign{{\cal F}_{0}^{(k)}(
S_i,\Lambda_1)= &{\Lambda_1^3\over12}(
S_1-S_2)+\sum_{j=1,2}{S_j^2\over2}\ln\Bigl({S_j
\over\Lambda_1^3}\Bigr)-{i\over 4}\kappa(S_1^2- S_2^2)\cr
-&{i\over2}\kappa S_1S_2- {3\over4}(S_1^2+S_2^2)
 +\Lambda_1^6\sum_{n\geq3}\sum_{j=0}^n c_{n,j} \Bigl({
S_1\over\Lambda_1^3}\Bigr)^{n-j}\Bigl({
S_2\over\Lambda_1^3}\Bigr)^j. }}
Note that in the present
derivation the $M_i$ are identified with the dimensionless
quantities $S_i/\Lambda_1^3$ and there is no need to add by hand
any additional scale. Furthermore, we note that one can
consistently define the matrix model with $M_i$ replaced by
${S_i/\Lambda_1^3}$. After this identification is made, one
analytically continues ${S_i/\Lambda_1^3}$ so that the critical
case can be consistently considered.

By \dacompararesoprabitod\ and \nuovaa\ the gauge theory effective
superpotential is
$$
W^{(k)}_{eff}(S_i)=\sum_{j=1,2}\Bigl({\partial{\cal
F}_0^{(k)}\over\partial S_j}-2\pi i \tau_0^{(k)}S_j\Bigr)=
$$
\eqn\odnmorteplici{ \sum_{j=1,2}S_j\left[\ln\Bigl({S_j\over
\Lambda_1^3}\Bigr)-1-4\ln\Bigl({\Lambda_2\over\Lambda_1}\Bigr)\right]
+i \kappa S_2 +\Lambda_1^3\sum_{n\geq2}
\sum_{j=0}^nd_{n,j}\Bigl({S_1\over\Lambda_1^3}\Bigr)^{n-j}
\Bigl({S_2\over\Lambda_1^3}\Bigr)^j,} where
$d_{n,j}=c_{n+1,j}(n+1-j)+c_{n+1,j+1}(j+1)$. Observe that
\eqn\relativoo{ W^{(k)}_{eff}(e^{2k_i\pi
i}S_i)=W^{(k)}_{eff}(S_i)+\sum_{j=1,2}2\pi i k_iS_i, } where
$k_i\in{\bf Z}$, $i=1,2$. It follows that the linear contribution
to the effective superpotential has the structure
$\sum_{j=1,2}(2\pi i k_j-1)S_j+ \pi i S_2$, so that the term $\pi i
S_2$ plays a special role as it cannot be completely reabsorbed by
a phase shift of the $S_i$.

Minimizing $W^{(k)}_{eff}$ in \odnmorteplici\ gives the two $F$--term conditions
\eqn\ancoratu{ S\equiv
S_1=e^{i \kappa}S_2,}
and
\eqn\oifww{ \Bigl({\Lambda_2\over\Lambda_1}\Bigr)^4=
{S\over\Lambda_1^3}\exp\Bigl[\sum_{n\geq3}b_n
\Bigl({S\over\Lambda_1^3}\Bigr)^{n-2}\Bigr]. }
In order to identify the
${\cal N}=2$ effective coupling constant, we should first
recognize the relationship between the two scales $\Lambda_1$
and $\Lambda_2$ at the extremum.

Additional contributions to the Veneziano--Yankielowicz superpotential \VenezianoAH\
have already been considered in literature, for example by
Kovner and Shifman \KovnerIM. More recently, Cachazo, Seiberg and Witten \CachazoZK\
first observed that in considering the critical points, which follow from the Veneziano--Yankielowicz superpotential
$W^{VY}$, one may consider, in the case of $SU(N)$, either $\ln({\Lambda^{3N}/S^N})=0$ or $N\ln({\Lambda^{3}/S})=0$,
leading to an apparent ambiguity. Then they observed that $W^{VY}$
can be defined on each of the $N$ possible infinite cover of the $S$--plane. In particular, according to
their analysis, one should explicitly include additional branches to $W^{VY}$. In the case of symmetry breaking
pattern $U(N)\to\prod_jU(N_j)$ they obtained
\eqn\csww{W_{eff}(S_i)=\sum_{j=1}^n \left(2\pi i \tau_0S_j + N_jS_j
 \left[\ln(\Lambda_j^3/ S_j) +1\right] + 2\pi ib_jS_j\right)
 + {\cal O}(S_i S_j),}
where the $b_j$ are integers with $b_1=0$. It is interesting to observe
that the additional contribution $2\pi i\sum_jb_jS_j$,
representing the relative shifts of theta vacua,
is reminiscent of the $i \kappa S_2$ term
in \odnmorteplici. However, note that whereas
${\kappa/\pi}$ is odd, the additional term in \csww\ always
depends on the even numbers $2b_j$. Furthermore, unlike \odnmorteplici,
for each $N_j=1$,
the corresponding term $2\pi ib_jS_j$ in \csww\ can be exactly obtained, as it should, by
the phase shift $S_j\to e^{2\pi i b_j}S_j$ in the argument of $W_{eff}$. The reason
is that the $b_j$'s label the theta vacua of each factor in the broken gauge group, and so
they play no role in the Abelian case.
Therefore, even if these contributions have a similar structure, they appear of different
nature. In particular, whereas the  Cachazo, Seiberg and Witten term is based on the
general properties of the logarithm, our additional term is a consequence of
the request that at the critical point $S_1+S_2=0$, as it should in the model we
are considering. Nevertheless, in spite of the differences, it is likely
that further investigation in this direction may lead to a better understanding of the Veneziano--Yankielowicz
superpotential and related issues.

\subsec{SW modulus and the scales}

Apparently, we do not have any information concerning the
identification of the $u$ modulus. In previous formulations this
was argued by identifying the parameters of the matrix model
potential and the SW curve. Here we have a different view which is
strictly related, as the structure of the bare coupling constant
$\tau_0^{(k)}$ indicates, to the RGE. In the previous approaches one
identified $u$ in terms of $\Delta$ using the derivation of the
matrix potential from the SW curve, leading to some questions
outlined in previous sections.
In our case, by \oifww\ and the nonperturbative relation \muuu, which
still holds for the free energy \dacompararesoprabitod, we recover again the identity
\edajjie, as explained in detail in \work. Therefore, by means of monodromy
arguments we can make the identification at the critical point
\eqn\ledueesse{ {{\cal
S}\over\Lambda_2^3}={S\over\Lambda_1^3}. } Furthermore, this
implies that the left hand side of \oifw\ and that of \oifww\
coincide, i.e.
$$
\left({\Lambda_2\over\Lambda_1}\right)^4=\left({\Lambda_2^2\over4u}\right)^2,
$$
by means of which we recover the relation between the $u$--modulus
and the ${\cal N}=1$ scale
\eqn\allestremo{ \Lambda_1^2=4 u, } and
\ledueesse\ becomes ${\cal
S}/\Lambda_2^3=S/8u^{3/2}$. Thus we have found
that in the present formulation at the critical point the ${\cal
N}=1$ scale coincides with $2$ times the square root of the
$u$--modulus of the ${\cal N}=2$ theory.

We can as well modify the prescription \ns\ by means of \ledueesse\ and make the following
identification in \preliminare\ \eqn\nsss{ M_1={{\cal
S}_1\over\Lambda_2^3},\qquad M_2={{\cal S}_2\over\Lambda_2^3}, }
with the free energy given by \eqn\lasoluzioneddd{ {\cal
F}_{0}^{(k)}( {\cal
S}_i,\Lambda_2)=\Lambda_2^6F_{0}^{(k)}\Bigl({{\cal
S}_i\over\Lambda_2^3}\Bigr). }
By evaluating ${\cal F}_{0}^{(k)}( {\cal S}_i,\Lambda_2)$ directly at ${\cal
S}\equiv {\cal S}_1=e^{i\kappa}{\cal S}_2$, we see that the
$k$--dependence completely disappears. In particular, we obtain \eqn\effeee{
{1\over4\pi i}{\cal F}_{0}^{(k)}( {\cal S},e^{-i\kappa}{\cal
S},\Lambda_2) ={\Lambda_2^3\over24\pi i}{\cal S} -{3\over8\pi
i}{\cal S}^2 +{1\over4\pi i}{\cal S}^2\ln{{\cal
S}\over\Lambda_2^3}+{\Lambda_2^6\over4\pi i
}\sum_{n\geq3}a_n\Bigl({{\cal S}\over\Lambda_2^3}\Bigr)^n,}
that now precisely corresponds to the SW prepotential as obtained
by integrating twice $\tau$ with respect to the glueball superfield.

Note that the absence of any parameter in the expression of the
free energy allows us to look for its monodromy properties. To
understand this aspect, observe that if the term
$(S_1^2+S_2^2)\ln\mu$ is present in the expression of the free
energy, the monodromy would involve $\mu$ dependent terms leading
to a rather involved analysis.

In the usual formulation the potential
depends on some parameters, namely the couplings,
whereas in \clamoroso\ they are missing. This is due to the fact
that simply we need not double the number of parameters.
Actually, once $S_1$, $S_2$ and ${\cal F}_0$ are given, we have
enough information to get the full SW theory. In particular,
the above discussion shows that the $u$--modulus arises in
terms of $a$ through the relation \muuu.

\newsec{Triangulating the Instanton Moduli Space}

Results in noncritical strings uncovered a deep connection between
algebraic--geometrical structure and Liouville theory. It should
be stressed that, on one hand, Liouville theory arises in the
description of the moduli space of Riemann surfaces, in particular
the Liouville action is the K\"ahler potential for the
Weil--Petterson metrics. On the other hand, Liouville theory is
the crucial quantum field theory for noncritical strings. In
particular, in \LG\ it was shown that there is an analytic
formulation for 2D pure quantum gravity which is directly
expressed in terms of the Liouville geometry of moduli space of
punctured spheres, reproducing the Painlev\'{e} I (Liouville
$F$--models). In that paper it was also argued that the
eigenvalues of the matrix model should be seen as punctures on a
Riemann sphere, which can be identified with the branch points on
the Riemann sphere itself. We note, in passing, that the relation
between punctured spheres and hyperellittic Riemann surfaces leads
to relations between Weil--Petterson volumes for such surfaces,
e.g. $Vol_{WP}(\overline{\cal M}_{1,1})=2Vol_{WP}(\overline{\cal M}_{0,4})$.
Furthermore, there is the isomorphism $\overline{{\cal
M}}_{2,0}\cong\overline{{\cal M}}_{0,6}$. One may expect that
these relationships hide more general properties of moduli spaces,
which should be strictly related to the Deligne--Knudsen--Mumford
compactification. The latter, together with the Wolpert
restriction phenomenon, is at the heart of the recursion relations
associated to the Painlev\'{e} I as derived in \LG. The analogy
with the recursion relation for the ${\cal N}=2$ instantons
suggested the formulation of instanton numbers in terms of
intersection theory \sfere. Recalling that 2D quantum gravity also
leads to a natural triangulation of moduli space of Riemann
surfaces, one might expect that a similar structure arises in
instanton theory. Remarkably, we have that the coincidences
discussed in Section 2 provide the following direct identification
of the $\N=2$ prepotential \eqn\tringa{ e^{{\hat{\cal
F}/\Lambda_2^6}}={1\over{\rm Vol}(U(M_1))\times{\rm
Vol}(U(M_2))}{\int {\cal D}\Phi_1{\cal D}\Phi_2
e^{-W^{(k)}(\Phi_1,\Phi_2)}}|_{M_1=e^{i\kappa}M_2\equiv {\cal
S}/\Lambda_2^3}, } where
$$
W^{(k)}(\Phi_1,\Phi_2)=W_1^{(k)}(\Phi_1)+W_2^{(k)}(\Phi_2)+W_I^{(k)}(\Phi_1,\Phi_2).
$$
This gives a direct way of expressing $\N=2$ instanton
contributions, including the gravitational corrections considered
in \NekrasovQD\FlumeAZ\BruzzoXF\KMT, in terms of a matrix model.
The above remarks then suggest that this formulation should be
related to a kind of triangulation of instanton moduli space.

\vskip 20pt

\noindent {\bf Acknowledgements}. The authors would like to thank
L. Alday, D. Bellisai, G. Bertoldi, G. Bonelli, M. Cirafici, R.
Dijkgraaf, F. Fucito, E. Gava, H. Ooguri, P. Pasti, G.C. Rossi, M. Serone, D.
Sorokin, M. Tonin, G. Travaglini and M. Tsulaia
for useful discussions. Work partially supported by the
European Community's Human Potential Programme under contract
HPRN-CT-2000-00131 Quantum Spacetime.

\listrefs

\end